# A survey of Hardware-based Control Flow Integrity (CFI)

RUAN DE CLERCQ* and INGRID VERBAUWHEDE, KU Leuven

Control Flow Integrity (CFI) is a computer security technique that detects runtime attacks by monitoring a program's branching behavior. This work presents a detailed analysis of the security policies enforced by 21 recent hardware-based CFI architectures. The goal is to evaluate the security, limitations, hardware cost, performance, and practicality of using these policies. We show that many architectures are not suitable for widespread adoption, since they have practical issues, such as relying on accurate control flow model (which is difficult to obtain) or they implement policies which provide only limited security.

CCS Concepts: • **Security and privacy** → **Hardware-based security protocols**; **Information flow control**; • **General and reference** → *Surveys and overviews*;

Additional Key Words and Phrases: control-flow integrity, control-flow hijacking, return oriented programming, shadow stack

## 1 INTRODUCTION

Today, a lot of software is written in memory unsafe languages, such as C and C++, which introduces memory corruption bugs. This makes software vulnerable to attack, since attackers exploit these bugs to make the software misbehave. Modern Operating Systems (OSs) and microprocessors are equipped with security mechanisms to protect against some classes of attacks. However, these mechanisms cannot defend against all attack classes. In particular, *Code Reuse Attacks (CRAs)*, which re-uses pre-existing software for malicious purposes, is an important threat that is difficult to protect against.

*Control Flow Integrity (CFI)* is a term used for computer security techniques which prevent CRAs by monitoring a program's flow of execution (*control flow*). CFI techniques do not aim to prevent the sources of attacks, but instead rely on monitoring a program at runtime to catch deviations from the normal behavior. CFI can detect a wide range of attacks, since no matter which attack vector is used, most attacks will eventually cause abnormal program behavior, such as an unexpected control flow change or the execution of tampered instructions.

---

*The corresponding author

This work was supported in part by the Research Council KU Leuven: C16/15/058. In addition, this work is supported in part by the Flemish Government through G.0130.13N and FWO G.0876.14N, and Cathedral ERC Advanced Grant 695305.

Author's addresses: R. de Clercq and I. Verbauwhede, imec-COSIC, KU Leuven, 3001-Leuven, Belgium.

CFI has received a lot of attention by the research community, but has not yet been widely adopted by industry. This could be due to the practical challenges and limitations of enforcing CFI, such as requiring complex binary analysis or transformations, introducing unwanted overhead, or offering incomplete protection. In this paper, we present an analysis of the security policies enforced by 21 state-of-the-art *hardware-based CFI* architectures. The scope of the paper is limited to works that introduce new hardware features/components to enforce CFI. Therefore, CFI architectures that do not introduce new hardware features/components are considered out of scope. As such, software-based solutions are only briefly discussed, and the reader is referred to [9] for a survey of software-based CFI. We identified a total of 13 security policies, and discuss each policy in terms of its security, limitations, hardware cost, and practicality for widespread deployment.

The remainder of the paper is structured as follows. First, we motivate the need for CFI by introducing the recent history of attacks and countermeasures. After that, we provide some background on the enforcement of CFI, the need for hardware-based CFI, together with a description of two methods for interfacing the CFI hardware monitor with the processor. Next, we introduce the three different kinds of attackers assumed by most CFI architectures. Subsequently, we present the classical approach for enforcing CFI together with its challenges and limitations. Afterward, we present the CFI policies used by the 21 hardware-based CFI architectures evaluated in this paper, followed by a discussion on CFI enforcement via the processor's debug interface. Finally, we provide a detailed comparison of the architectures and their hardware costs/overhead, followed by a conclusion.

## 2 ATTACKS AND COUNTERMEASURES: AN ARMS-RACE

This section provides a brief history of attacks and countermeasures.

We define a *memory error* as a software bug which is caused by invalid pointer operations, use of uninitialized variables, and memory leaks. Of particular importance is memory corruption, which typically occurs when a program unintentionally overwrites the contents of a memory location. Memory errors are produced by memory unsafe languages, such as C and C++, which trade type safety and memory safety for performance. These errors are prevalent in a large amount of deployed software: (1) memory unsafe languages are frequently used to write OS kernels and libraries which are also used by memory safe languages, while (2) some memory safe languages, such as Java, rely on an interpreter which is written in a memory unsafe language. Attackers exploit these memory errors to intentionally overwrite a memory location. Defending against the exploitation of existing software bugs has led to an arms-race between attackers and defenders, where each new defense leads to an attack which circumvents it.

An important example of exploiting a memory error is the *buffer overflow*, where more data is written to a buffer than the allocated size, leading to the overwrite of adjacent memory locations. Attacks frequently make use of a buffer overflow by providing input data that is larger than the size of the buffer. This causes other items to be overwritten, such as local variables, pointer addresses, return addresses, and other data structures.

Many *code injection attacks* rely on a stack-based buffer overflow to inject shellcode onto the stack and overwrite the return address. By overwriting the return address, the attacker can change the control flow to any location during a function return. The attacker uses this to divert control flow to the injected shellcode, thereby allowing him to execute arbitrary code with the same privileges as that of the program. To defend against return address tampering, *stack canaries* [7, 17] involve placing a canary value between the return address and the local function variables. The canary value is verified by a code fragment before returning from the function. However, canaries have been shown to be vulnerable [5] and further require the insertion and execution of additional instructions at the end of each function call.

Another attack vector, known as *format string vulnerabilities* [44], allow overwriting arbitrary addresses. This vector can be used to overwrite return addresses without changing the canary value.

An effective defense against code injection attacks is non-executable (NX) memory [1], a.k.a W ⊕ X (Write XOR eXecute), a.k.a Data Execution Prevention (DEP). An NX bit is assigned to each page to mark it as either readable and executable, or non-executable but writable. Most high-end modern processors have architectural support for W ⊕ X , while most low-end processors do not. This protection mechanism was circumvented by the invention of *code reuse attacks (CRAs)*, which do not require any code to be injected, but instead uses the existing software for malicious purposes. An example of this is the *return-to-libc* attack, where the attacker updates the return address to force the currently executing function to return into an attacker-chosen routine. In addition, the attacker places function arguments on the stack, thereby providing him with the ability to supply attacker chosen arguments to a function. While the attacker could return anywhere, libc is convenient since it is included in most C programs. A popular libc attack target is to spawn a shell by returning into `system("/bin/sh")`, or to disable W ⊕ X by returning into `mprotect()`/`VirtualProtect()`.

*Return oriented programming (ROP)* [49] is a powerful CRA which is Turing-complete[1]. ROP makes use of code *gadgets* present inside the existing program. Each gadget consists of a code fragment that ends with a return instruction. The attacker overwrites the stack with a carefully constructed sequence of gadget arguments and gadget addresses. The goal is to invoke a chain of gadgets, with each return instruction leading to the invocation of the next gadget. After the initial function return, the first gadget is executed, leading to the eventual execution of a return, which causes the next gadget to be executed.

Another type of attack, called *Jump oriented programming (JOP)* [8], is also Turing-complete. The building blocks are also called gadgets, but here each gadget ends with an indirect branch instead of a `ret` instruction. A dispatch table is used to hold gadget addresses and data, and a CPU register acts as a virtual program counter which points into the dispatch table. A special gadget, called a dispatcher gadget, is used to execute the gadgets inside the dispatch table in sequence. After invoking each functional gadget, the dispatcher gadget is invoked, which advances the virtual program counter and then launches the next functional gadget.

*Code randomization* protects against CRAs by placing the base addresses of various segments (`.text`, `.data`, `.bss`, etc) at randomized memory addresses. This makes it difficult for attackers to predict target addresses, and is currently used in one form or another in most modern OSs. The security of this technique relies on the low probability of an attacker guessing the randomly placed areas. Therefore, a larger search space means more effective security. The Linux PaX project introduced a code randomization technique, called Address Space Layout Randomization (ASLR), with a patch for the linux kernel in July 2001. ASLR suffers from two main problems. First, the effectiveness of this technique is limited on 32-bit architectures due to the low number of bits available for randomization, which makes the system vulnerable to brute force attacks [50]. Second, it is vulnerable to memory disclosure attacks, since only the base addresses of each segment is randomized. If an attacker gains knowledge of a single address, he could compute the segment base address, which causes the system to again become vulnerable to CRAs. A method to disclose a memory address is by exploiting a format string vulnerability.

*Non-control data attacks* [13] rely on corrupting data memory which is not directly used by control flow transfer instructions. In the past, it was assumed that non-control data attacks were limited to data leakage (e.g., HeartBleed [22]) or data corruption. However, recently this attack vector was used to launch two different Turing-complete attacks which can circumvent CFI policies (see Section 5.1.1 and Section 5.3.3).

---

[1]A Turing-complete language can solve any possible computation problem.

## 3   BACKGROUND

### 3.1   Control Flow Integrity (CFI)

The goal of CFI is to detect runtime attacks by monitoring program behavior for abnormalities. A *Control Flow Graph (CFG)*, which is a model of the valid control flow changes inside a program, is commonly used to model normal program behavior. Each node in the CFG represents a *basic block*, which is a group of instructions where control strictly flows sequentially from the first instruction to the last. Therefore, a basic block has a single entry point at the first instruction, and one exit point at the last instruction, which is the only instruction that may cause a change in control flow. In a CFG, *forward edges* are caused by jumps and calls, while *backward edges* are caused by returns. The CFG is typically generated by statically analyzing the source code or the binary of a program.

At runtime, the dynamic control flow changes are restricted to the static CFG. Many CFI architectures only validate control flow changes caused by indirect branches, such as calculated jumps, calculated calls, and returns. The assumption is that the software is immutable, which means that static branch targets do not need to be checked. This implies that this policy cannot be used for self-modifying code, or for code generated just-in time. Software-based CFI policies typically verify each indirect branch target before the execution of an indirect branch instruction.

We define *fine-grained* CFI as a policy which only allows control flow along valid edges of a CFG. In recent years, many works proposed architectures which relax the strict enforcement of the CFG, in order to gain performance improvements. We define *coarse-grained* CFI as a policy which does not enforce a strict CFG. These policies rely on enforcing simple rules, such as ensuring that return targets are preceded by a `call` instruction, and that indirect branches can only target the first address in a function. They offer less security than fine-grained CFI policies, as demonstrated by recent attacks [20, 26].

### 3.2   The need for hardware-based CFI

Many software-based CFI solutions rely on instrumentation, where code is inserted into a program to perform CFI checks on indirect branches. This can be done as part of a compiler optimization step, static binary rewriting, or through dynamic library translation. When the compiler is unaware of the security aspects concerning the CFI checks, it might cause the optimization step to spill registers holding sensitive CFI data to the stack. Static metadata is often protected with read-only pages. However, many software-based CFI architectures rely on runtime data structures which are stored in memory which is sometimes writable. Recent attacks [16, 23] exploited this problem by tampering with runtime data structures to circumvent the security of the system. However, runtime data structures can be protected through instruction-level isolation techniques, such as memory segmentation or Software-based Fault Isolation (SFI). Memory segmentation provides effective isolation on x86-32, but is not really useful on x86-64 since segment limits are not enforced. Software-based Fault Isolation (SFI) executes additional guards to ensure that each store instruction can only write into a specific memory region. It is worth pointing out that instruction-level isolation techniques rely on the integrity of code. Hardware-based access control mechanisms can provide strong isolation for runtime data structures and metadata. In addition, it has little overhead, and does not rely on code integrity to protect sensitive information.

The *Trusted Computing Base (TCB)* is the set of hardware and software components which are critical to the security of the system. The careful design and implementation of these components are paramount to the overall security of the system. The components of the TCB are designed so that other parts of the system cannot make the device misbehave when they are exploited. Ideally, a TCB should be as small as possible in order to guarantee its correctness [40].

Software-based CFI uses instrumented code to monitor the behavior of other software. To prevent tampering, the software is protected by page-based access control (such as W ⊕ X ). However, this only temporarily makes the software immutable, and care needs to be taken to ensure that an attacker cannot disable page-based protection with a syscall to `mprotect()`/`VirtualProtect()`. In contrast, the underlying hardware architecture is immutable, and is therefore more trusted. To enforce a strong security policy, we recommend that the TCB consists of as little as possible software, while placing as much as possible security-critical functionality in hardware.

Hardware-based mechanisms can protect against strong attackers, such as attackers which control both code and data memory, as well as attackers with physical access that can perform fault attacks. In addition, it allows for verifying not only indirect branches, but also for direct branches, unintended branches, and for control flow between instructions inside a basic block, which may be altered during a fault attack. Furthermore, the TCB size is usually smaller, since security-critical functionality is mostly implemented in hardware, with littler or no trusted software. Moreover, hardware-based mechanisms provide better performance, since fewer processor cycles need to be spent on performing security checks.

However, hardware-based CFI is no silver bullet. An important design decision for security-critical systems is to select which components need to be placed inside the TCB. Many hardware-based CFI architectures need software components to communicate runtime information with the CFI hardware or configure the CFI hardware. Care needs to be taken to ensure: the correctness of the TCB components, that the hardware/software interface cannot be exploited, that sensitive data which is stored in main memory is protected, and that the hardware-based security policies cannot be circumvented.

### 3.3 Hardware monitor

Hardware-based CFI architectures use trusted hardware to monitor the behavior of the software, which we call the *hardware monitor*. Providing the hardware monitor with access to the necessary runtime information and signals, such as control flow information, runtime data structures, and metadata is an important issue when designing the CFI architecture. To access runtime control flow information, the following two approaches can be used: (1) integrating the hardware monitor into a processor by modifying the instruction pipeline stages, and (2) interfacing the hardware monitor with the processor's hardware debugger. To access runtime data structures or metadata stored in main memory, the hardware monitor can be connected as a master to the processor's bus, thereby providing access to the processor's memory space.

*3.3.1 Processor's pipeline stage.* Many architectures integrate their hardware monitor into one or more of the processor's instruction pipeline stages, as shown in Fig. 1. This allows all the processor's internal signals to be exploited for enforcing the CFI policy. For example, many CFI policies require runtime branch information such as the branch source address, target address, and branch type. To enforce CFI, the required signals can be forwarded from inside the appropriate pipeline stage to the hardware monitor, which uses this information to verify the validity of each branch.

Another approach is to add *special instructions* to the ISA to enforce a CFI policy. This also requires modification of pipeline stage(s), such as the Instruction Decode (ID) stage. In order to make use of the special instructions, a software toolchain is often used to insert these special instruction into the compiled software.

In Fig. 1 the hardware monitor is integrated with the "ID" pipeline stage. However, the monitor can be integrated into multiple different pipeline stages. The decision on where to integrate the monitor depends on the availability of information at a given stage.

See Section 6 for a presentation on the state-of-the-art in CFI architectures that are integrated into the pipeline stages of the processor.

The biggest disadvantage of integrating a monitor into the pipeline stages is that it does not follow the typical System-on-Chip (SoC) design flow, as it requires modification of existing Intellectual Property (IP), such as the processor.

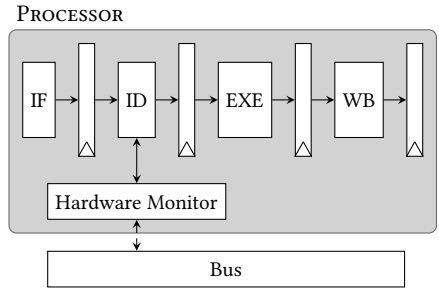

Fig. 1. The hardware monitor is integrated into the instruction pipeline of the processor.

*3.3.2 Debug interface.* Some recent works exploit the processor hardware debug interface to monitor a program's behavior without modifying the processor or other pre-existing IP, as shown in Fig. 2. This is a major advantage, as it complies with the design rules of SoCs. Modifying IP is expensive, as it requires modification by the IP vendor (expensive), or requires the IP vendor to release its Hardware Description Language (HDL) to the client (even more expensive). Monitors which are realized by hardware-based standalone IP are practical, because they do not require the modification of the target processor, but simply communicate with the target processor via existing interfaces, such as the processor's hardware debug port or the bus.

A major challenge with enforcing CFI through the debug interface, is to obtain access to all the required runtime information. Since the hardware debugger is designed for debugging purposes, it usually does not provide all the necessary information to enforce a given protection mechanism. A common approach to solve this is to provide supplementary metadata inside main memory, or to communicate the missing information by instrumenting the binary to write the required information to memory-mapped addresses (MMIO).

Some SoCs have an external debug interface, which facilitates the use of an external hardware monitor. In this case the monitor can be realized by an off-chip microprocessor or other dedicated hardware (FPGA or ASIC), which are connected to the external debug interface.

See Section 7 for a presentation on the state-of-the-art in CFI architectures which monitor the processor via the hardware debug interface.

## 4   ATTACKER MODEL

The goal of the CFI architecture is to prevent attackers from launching a CRA. In contrast, the goal of the attacker is to use a CRA to perform arbitrary code execution, confined code execution, arbitrary computation, or information leakage.

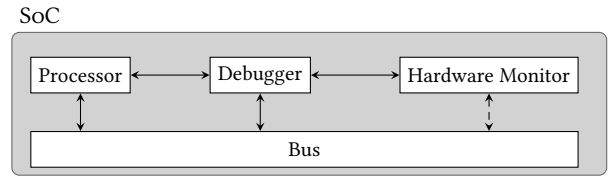

Fig. 2. A CFI hardware monitor interfaces with the processor's debug port. In addition, a memory mapped interface (MMIO) is also used by instrumented code to communicate with the hardware monitor.

In general, the architectures described in this paper use one of three different attacker models. However, each of the three attacker models share the following assumptions.

- The memory contents and layout are readable by the attacker. The usual assumption is that ASLR can be circumvented, and that the location of all program segments are known.
- The hardware is trusted.
- Memory errors can be present inside the program. However, the usual assumption is that no other attack vectors or security holes exist which could directly allow an attacker to perform a privilege escalation.
- The attacker can arbitrarily write to memory by exploiting a memory error. The exact locations of writable memory depend on the used attacker model.
- Side-channel attacks are not possible.

The *first attacker model* assumes that the attacker is in control of data memory (which includes the stack and the heap), but not code memory. The usual assumption is that W ⊕ X protection is enforced, which means that code injection and code tampering attacks cannot be performed. This further means that the OS is implicitly trusted, since OS support is required to set the page bits to enforce W ⊕ X . In order to launch an attack, the attacker has to exploit a memory error to overwrite data memory, such as code pointers or other data structures.

The *second attacker model* assumes that the attacker is in control of data memory as well as code memory. An attacker can therefore perform code injection and code tampering, since W ⊕ X is not enforced. However, architectures using this attacker model usually enforce a *Software Integrity (SI)* mechanism to prevent the execution of tampered/injected code. In this model, an attack can be performed by exploiting a memory error to overwrite any memory location.

The *third attacker model* also assumes an attacker in control of all memory. In addition, the attacker can perform non-invasive fault attacks that target the program flow, such as glitching the clock. However, other types of fault attacks that do not target the program flow, such as glitching the ALU or bus of the processor, are not considered part of the attacker model. Therefore, in this scenario the attacker has two avenues for launching an attack: exploiting a memory error to overwrite any memory location, or performing a fault attack on the control flow.

## 5 CLASSICAL CFI

This section discusses Abadi et al.'s [4] classical approach for enforcing CFI. Even though this policy was originally enforced in software, many of its challenges and limitations are also applicable to hardware-based CFI. For an in-depth discussion on hardware-based CFI policies, see Section 6.

### 5.1 Labels

The label-based approach relies on inserting unique label IDs at the beginning of each basic block. Before each indirect branch, a the destination basic block's label ID is compared to a label ID which is stored inside the program. In other words, the correctness of the destination basic block's label ID is verified before every indirect branch. Since unique label IDs are used, control flow tampering causes the check to fail, since the destination label ID will not match the label ID stored inside the program. The control flow checks are performed using code checks which are inserted at the end of each basic block containing an indirect branch.

*5.1.1 Limitations.* A *static CFI* policy (such as the label-based approach of [4]), can only check that control flows along the CFG. It is *stateless*, since the the stack's runtime state is not considered when determining which paths on the CFG are valid. This means that it cannot be used to ensure that a function returns to its most recent call site. In other words, when enforcing a static CFG, each function can return along any valid backward edge inside the CFG. This is a problem for any function which is called from more than one location. During normal execution, the function might be called from one location, but after an attacker overwrites the return address, the function can return along another CFG edge to a different valid call site (of the attacker's choosing).

Carlini et al. [10] demonstrated the severity of the sateless problem, by performing a non-trivial attack in the presence of a strict static CFI policy. A *Control-Flow Bending (CFB)* attack relies on *bending* control flow along valid CFG paths by overwriting return addresses. This allows an attacker to rapidly traverse the CFG in order to execute any attacker-chosen code.

### 5.2 Shadow Call Stack (SCS)

In order to overcome the stateless problem of static CFI and enforce a strict backward-edge policy, a *Shadow Call Stack (SCS)* can be used. The goal of an SCS is to detect tampering with return addresses stored on the stack during a function call. To achieve this, an additional stack is used to store an extra copy of the return address for each function call. Therefore, every `call` requires that two copies of the return address are stored, one inside the stack frame, and the other inside the SCS. Before returning (`ret`) from a function call, the integrity of the return address is verified by comparing the return address stored on the two stacks. If there is a mismatch, an exception is raised. In essence, this means that the SCS ensures that every call to a function returns to its call site.

A major advantage of enforcing an SCS is that it provides an excellent defense against attacks on backward edges, which include ROP and return-into-libc attacks. In addition, it doesn't rely on a CFG, which is problematic to obtain in complex binaries. These factors make hardware-based SCS suitable for widespread adoption.

When a stateful backward-edge policy, such as SCS, is used together with a forward-edge CFI policy (such as the label-based approach), we call it *dynamic CFI.*

### 5.3 Challenges and limitations

*5.3.1 Generating a precise CFG.* Many CFI policies determine the set of valid branch targets by calculating a CFG of the program through static analysis. However, generating a precise CFG through static analysis is an unsolved problem. The problem is that it is difficult to compute a precise list of valid branch targets for all indirect branches. To solve this, over-approximation is used, leading to a CFG that contains more edges than necessary, and is therefore not fully precise [10, 34]. This degrades the security of the CFI scheme relying on the CFG, since the security of CFI depends on an accurate CFG. For a classification of the precision of computing a CFG using different static analysis techniques, the reader is referred to [9].

Therefore, security architectures which rely on a precise CFG cannot enforce a strict policy, since a precise CFG cannot be generated for indirect branches. However, if a program contains no indirect forward branches, then this is not a problem, since static analysis can accurately generate a CFG for direct branches and function returns. For classical CFI this is not useful, since it is only concerned with enforcing a CFG for indirect branches.

*5.3.2 Unintended branches.* Unintended branches can occur in architectures which use a variable length instruction encoding. To exploit it, the attacker deviates control flow to the middle of an instruction. Many coarse-grained policies cannot check for unintended branches. This is a serious concern, since the majority of gadgets in a program consists of unintended branches, e.g., 80% of all libc gadgets are due to unintended branches [32].

*5.3.3 Limitations.* Here we discuss some advanced attacks that can circumvent the security of dynamic CFI.

A *printf-oriented programming* attack [10] provides Turing-complete computation, even when enforcing a strict CFI policy together with a SCS. Here, the attacker exploits the standard library's printf function by controlling the format string, arguments, and destination buffer. The printf function allows for performing arbitrary reads and writes, and conditional execution. In addition, the attack is Turing-complete when a loop in the CFG can be exploited.

*Counterfeit Object Oriented Programming (COOP)* [48] demonstrated that C++ virtual functions can be chained together to create Turing-complete programs even when precise dynamic CFI is enforced. The attack injects counterfeit C++ objects into the program address space followed by a manipulation of virtual table pointers. It is believed that this attack was possible because most CFI defenses do not considering C++ semantics.

*Data-oriented programming (DOP)* [29] can achieve Turing-complete computation by utilizing non-control data attacks in the presence of a precise and dynamic CFI policy. Similar to ROP, these attacks utilize data-oriented gadgets, which consists of short instruction sequences that enable specific operations. In addition, a dispatcher gadget is used to chain together a set of gadgets to force a program to carry out a computation of an adversary's choice. They demonstrated the attack by disabling page-based protection in the presence of a software-based dynamic CFI policy.

## 6 HARDWARE-BASED CFI POLICIES

This section presents the CFI policies used by the hardware-based CFI architectures evaluated in this paper. We discuss the benefits, challenges, and hardware cost for enforcing each policy in hardware.

### 6.1 Shadow Call Stack (SCS)

An SCS, as introduced in Section 5.2, is used by many hardware-based CFI solutions [6, 15, 18, 24, 28, 30, 32, 33, 36–38, 42, 51] to enforce a backward-edge policy, together with a range of different forward-edge policies. A hardware-based SCS typically requires the following components: a hardware buffer to store the most recently used entries, logic to manage the entries in the buffer, logic to interface with the main memory, and logic to handle exceptions (setjmp/longjmp).

*6.1.1 Hardware buffers.* When enforcing an SCS with a hardware monitor which has fast access to the on-chip main memory of a small embedded processor, it is possible to store the entire call stack without paying a performance penalty [24]. However, doing this for processors using off-chip DDR memory will incur a larger overhead, since each main memory access may take many cycles. Therefore, to reduce the number of main memory accesses, a hardware buffer is used to store the most recently used entries of the shadow stack. This can lead to a significant speedup, since accessing the hardware buffer can be done in a single cycle.

The next challenge is to select the buffer size. Since on-chip memory is expensive, the hardware buffer can only accommodate a limited number of entries. One approach is to design the buffer size around the maximum call depth of a program. However, [45] found a program in their benchmark suite has a maximum call depth of 238. In addition, it is expected that the maximum number of shadow stack entries will further increase with multithreading. Therefore, in order to keep the hardware cost low by having only a small buffer, and further allow for deeply nested calls, the older entries in the shadow stack can be stored in main memory. One approach [45] is to use an interrupt service routine to write the contents of the SCS to main memory when an overflow occurs. Another approach is to use a hardware-based *shadow stack manager* [39] to copy the oldest 50% of buffer elements to main memory upon detecing a buffer overflow. Most hardware architectures seem to use a buffer that accommodates a call depth of 16 or 32 entries.

*6.1.2    Protecting runtime data.*  A shadow stack stored in main memory allows for an almost unrestricted call depth. However, since we assume an attacker which controls data memory, a mechanism is needed to prevent tampering of runtime SCS data stored in main memory. IBMAC [24], which is implemented on a small microprocessor, uses an SCS memory region that can only be updated by `call` and `ret` instructions. Intel CET [30] uses special "SCS" pages such that regular store instructions cannot modify the contents of the shadow stack. Lee et al. [36] uses a hardware component that prevents the processor from accessing the SCS memory region by snooping traffic between the processor and the memory controller. HCFI [15] stores its entire call stack in a hardware buffer, which cannot be accessed by the processor. However, this restrictive approach can only accommodate a limited call depth. Some hardware-based approaches [18, 31, 32, 51] rely on memory allocated in kernel space to ensure that the SCS is separated from the process's memory space. Here, the OS kernel maps a region of physical memory and reserves it as part of MMIO. The pages can be marked as write-protected while still being updatable from the SCS hardware. Kernel memory is not accessible from user space, thereby only allowing the hardware monitor and the kernel to access the memory.

*6.1.3    Support for setjmp/longjmp.*  Complex binaries sometimes have exceptions to the normal behavior of `call`, `ret`, and `jmp`. One such case is `longjmp()`, which performs stack unwinding to restore the stack to the state it had during the previous `setjmp()` invocation. Therefore, after a `longjmp()`, when the subsequent return is made, the expected return address will not be on the top of the SCS, because it hasn't been unwinded yet. Smashguard [45] proposed to remove elements from the top of the stack until the top SCS element matches the return address.

HCFI [15] proposed to record the current SCS index when `setjmp()` is invoked, and unwinding the SCS to the recorded index for the subsequent `longjmp()` invocation. To support multiple `setjmp/longjmp` pairs, a unique 8-bit ID is assigned to each pair, and a 256-element hardware buffer is used to record the SCS indices for each unique ID.

Das et al. [18] noted that `longjmp()` sometimes uses an indirect `jmp` instruction instead of a `ret`. This causes an exception for the next executed `ret`, since the expected `ret` did not execute, and the top SCS element was not evicted. In addition, the compiler sometimes uses a `pop` and `jmp` pair instead of a `ret` instruction. To address these exceptional cases, Das et al. [18] proposed extended SCS rules. Whenever an indirect `jmp` targets an address inside the SCS, the corresponding address is removed from the SCS. BBB-CFI [28] proposed to improve the security of this approach by limiting `ret` targets to the top 8 elements of the SCS. This causes the matching element as well as the elements above it to be removed from the SCS. Special care needs to be taken when using this approach in combination with a fine-grained CFI policy, since an attacker could exploit this feature with a `ret` to any of the top 8 elements on the call stack.

BBB-CFI [18, 28] further noted that during software exceptions, the stack unwinding process is started, which frequently leads to a `ret` being used instead of a `jmp` to branch to an exception handler. This causes a problem when using an SCS, since the target return address will not be on the SCS. To solve this, they proposed to allow `ret` to target any of the exception landing pad addresses. In x86 ELF binaries the exception information are stored in the read-only `eh_frame` and `gcc_except_table` sections.

*6.1.4    Other considerations.* Recursive functions can be problematic, since a invoking a function many times can lead to a large memory overhead. HCFI [15] supports recursive functions by assigning a one-bit recursion flag to each SCS element. Before pushing a new return address onto the SCS, the address is first compared to the top element of the SCS. If the value is different, it is pushed. If the value is the same, then the top element's recursion flag is set. For each `ret`, the recursion flag remain set while the top element and the return addresses are the same. However, if there is a mismatch, the top element of the SCS is popped. This approach has the cost of storing one additional bit for every SCS element, which includes the hardware buffer and the storage in main memory. In [15], this was done using a 128*1-bit buffer, which restricts the call depth to 128.

SCSs used in a multithreading environment can also be problematic, since concurrency allows the `ret` of one thread to occur after the `call` of another thread, leading to SCS inconsistencies. Das et al. [18] proposed to relax the CFI policy by allowing returns to target any address stored in the SCS. This may reduce the security, since an attacker could re-order the sequence of returns. Using a Content-Addressable Memory (CAM) for storing the top SCS elements could accelerate searches. However, searching the SCS elements stored in main memory will incur a large performance penalty, since accessing main memory is slow. In the worst case the entire call stack will need to be searched.

A more strict alternative [18] is to store a process identifier (PID) in each SCS entry. Each `ret` will only use the return address associated with its specific PID. This approach has the cost of storing the PID inside every element of the call stack, which is stored in the on-chip hardware buffer and the off-chip main memory. (A PID requires $\geq$ 16 bits on Linux.) In addition, the call stack will need to be searched for a specific PID during each return. Searches can be accelerate with a small CAM containing one element for each PID. If the PID is not found on the CAM, a huge performance overhead will be incurred, since the buffer and main memory will need to be searched.

## 6.2    HAFIX: Shadow stack alternative

HAFIX [19] proposed a stateful, fine-grained backward-edge policy which does not rely on an SCS. To keep track of valid returns, a unique label is assigned to each function and a one-bit table entry is used for each label. During a function call the calling function's label is activated, while a return leads to the deactivation of the target function's label. During a return, if the target function's label is inactive, an exception is raised. To enforce this policy, HAFIX used special instructions inserted in the binary to activate/deactivate the labels. The ISA was modified to only allow function returns to a special landing pad instruction which performs the check. Recursive function calls are supported by counting the number of times a function has been called. This policy claims to improve upon an SCS by having a reduced memory overhead.

HAFIX is integrated into the processor's pipeline stages. The major hardware components of the design is the CFI control logic and the label state memory, which consists of a hardware buffer with a one-bit entry for each function.

### 6.3 Labels

HCFI [15] and Sullivan et al. [51] implemented Abadi's [3] label-based approach (see Section 5.1) in hardware through special instructions added to the ISA. Specifically, `SetLabel` is executed before each indirect forward branch to assign a new label ID to the label register, while the `CheckLabel` instruction is executed as the first instruction after an indirect forward branch, and is responsible for verifying the label ID stored in the label register [2]. Indirect branch targets are restricted to addresses containing the `CheckLabel` instruction.

It is important to highlight a security problem of the label-based policy. When multiple different indirect forward branches target the same basic block, the same label ID will be used, since `CheckLabel` compares the label register ID to a constant value. This means that all indirect forward branches targeting the same basic block need to assign the same label ID before branching. To make this problem worse, an indirect branch can target many different basic blocks, which all need to use the same label ID, since it is assigned before the branch. This causes imprecision in the enforced CFG, since the same label ID is used for multiple basic blocks, thereby allowing many more edges than in the original CFG. This problem can be prevented by using a *trampoline* for each basic block targeted by multiple call sites [51]. The code is transformed so that each indirect branch targets a unique trampoline. Each trampoline executes `CheckLabel`, and then performs a direct jump to the original target. This ensures that the precision of the enforced CFG is not reduced by large equivalence classes.

The designs of [15] and [51] are both integrated into the processor's pipeline stages. The major hardware components used in their designs are an SCS module, a label register, and logic for each special instruction.

### 6.4 Table

The *table-based* approach uses a table of allowed branches (source addr, dest addr), with a single entry for each direct branch, and possibly multiple entries for each indirect branch. At runtime, each branch is verified by checking for the existence of an entry inside the branch table.

This approach enforces fine-grained, static CFI, and requires a CFG to generate the branch table. The branch table could have a large storage requirement, and searching the table could incur a high performance penalty.

CONVERSE [27] enforced the table-based approach with a watchdog processor. It populates the branch table using both direct and indirect branches, by means of static and dynamic analysis. The architecture uses a watchdog processor to check the validity of each branch at runtime, as explained in detail in Section 7.1. The branch table is protected from attacks on the target processor, since it is stored in the memory space of the watchdog processor.

Arora et al. [6] proposed an architecture to enforce CFI and SI at runtime. Intra-procedural control flow is protected using the table-based approach, while inter-procedural control flow is protected using the *Finite State Machine (FSM)* approach (see Section 6.5). A hash is used to verify the integrity of the instructions in each basic block at runtime, and the processor is stalled to allow for completing the hash calculation before moving on to the next block. A program is loaded by first verifying the program and integrity of the metadata. Next, the hardware monitor's FSMs and tables are populated using the metadata, followed by executing the program. The authors claim that the architecture can eliminate a wide range of software and physical attacks. It seems that it can defend against physical attacks on control flow and the contents of basic blocks.

---

[2]For simplicity we used the instruction names `SetLabel` and `CheckLabel`, but HCFI named their instructions `SetPCLabel` and `CheckLabel`, while Sullivan et al. named their instructions `cfiprj`/`cfiprc` and `cfichk`.

 

The design of Arora et al. was integrated into the pipeline stages of the processor. The main hardware components are FSMs, lookup tables, buffers, a hash engine, and control logic. New FSMs and lookup tables are generated for each program, which means that each program has a different hardware area requirement. Inter-procedural checking can be implemented with a maximum area overhead of 1.2%, while the area increases to 5.17% when also adding intra-procedural checking, when compared to the die size of a ARM920T processor. When also using a hash engine, the area overhead increases to 9.1%.

### 6.5 Finite State Machine (FSM)

FSMs can be used to enforce a valid sequence of events. While a program is executing, a state machine tracks the events, with each event representing a node in the FSM. As long as the events inside an executing program follows the correct sequence, the state transitions will be valid. However, if an invalid state transition occurs, an attack is assumed, and appropriate action is taken.

Arora et al. [6] used an FSM to detect invalid control flow between functions. Each node in the FSM represents a function, and each edge represents control flow between functions. A function call graph is extracted from a program, with each function call or return corresponds to a state transition. For more information on this architecture, see Section 6.4.

Rahmatian et al. [46] used an FSM to detect invalid system call sequences. The observation is that malicious code must invoke system calls to perform some of the necessary operations to launch an attack. If the sequence of system calls deviates from the FSM, an attack is assumed. A model of the legal sequences of syscalls is derived from a CFG, and the policy is enforced with an FPGA. To detect a syscall, the architecture made modifications in the pipeline of the processor to detect a trap instruction (SPARCv8). The authors claim that the solution has zero performance overhead, and illegal system call sequences can be detected within three processor cycles. The system relies on software to update the FSM store on the FPGA at runtime. The major hardware component of the design is the syscall sequence FSM, which includes two 36Kb BRAM modules. A custom FSM is generated by static analysis, leading to different hardware requirements for different programs.

### 6.6 Heuristics

*Heuristic-based* approaches detect CRAs by deriving a CRA signature from the branching behavior of a program under attack. A popular CRA signature is the *number of executed instructions* between branches. The assumption is that typical ROP and JOP gadgets usually consist of a small number of instructions (around five instructions). Larger gadgets usually have side effects, such as updates to memory and registers, and are therefore avoided by attackers. A typical attack executes a number of short instruction sequences that each end with a branch, whereas a normal program executes larger instruction sequences between branches. Chen et al. [12] reported that ROP attacks require at least three gadgets, while Kayaalp et al. [33] reported that JOP attacks require at least six gadgets. This approach requires no CFG.

A major challenge is the proper selection of heuristic parameters (number of instructions vs chain length) to minimize the number of false positives. The number of false positives typically increase as the gadget length is incremented or when the chain length is reduced. In addition, the number of false positives can differ between different programs, while using the same parameter set.

Recent attacks [11, 20, 25] demonstrated that heuristics which only assume short gadgets can be circumvented. Detection can be avoided by placing an intermediate gadget (a gadget which doesn't do anything but is longer than the threshold of the heuristic) in the gadget chain. This is non-trivial, since the side effect(s) of the long gadget needs to either be tolerated, or subsequent gadgets should be used to undo the side effect.

To defend against intermediate gadgets, SCRAP [33] proposed a *multi-threshold detector*, which tolerates longer gadgets in the gadget chain. This is done by not advancing the gadget count when detecting an intermediate gadget. They demonstrated that they can protect the entire SPEC 2006 benchmark suite without any false positives. In addition, SCRAP limits false positives by allowing the chain length and instruction count to be configured at runtime. SCRAP [33] is integrated into the commit pipeline stage of the processor. The major hardware components of this architecture is an SCS module and the multi-threshold heuristic logic which consists of several state machine counters. A simulation using PTLsim/x86 showed a performance overhead results of around 2%.

The number of *consecutive indirect branches* can also be used as a CRA signature. The observation is that a well-behaved program executes a mix of direct branches as well as indirect branches, since a normal program usually contains many more direct branches than indirect branches. Therefore, to detect an attack, the heuristic counts the number of consecutive indirect branches. If more than $\gamma$ indirect branches are consecutively executed, then the program is assumed to be under attack. While this rule can detect most CRAs, adversaries can circumvent it by using a *gadget gluing attack* [14]. Here, the attacker places a special gadget containing a direct branch inside the gadget chain to thwart the heuristic. To address this issue, [39] proposed to use a second threshold parameter $\delta$, which represents the maximum number of direct branches that are allowed between indirect branches. Therefore, an alarm is only raised if the branch trace contains more than $\gamma$ indirect branches and at most $\delta$ direct branches.

Lee et al. [39] proposed a *two-stage heuristic* policy, where a mixed hardware/software approach is used for detecting CRAs in a resource-constrained environment. In the first stage, a lightweight hardware monitor employs a multi-threshold heuristic based on the number of consecutive indirect branches. When the first stage detects anomalous behavior, an interrupt is raised, and the second stage software performs an in-depth analysis on the occurred branching behavior, in order to ensure that the detected operation is not a false alarm. The idea of the two-step approach is to increase the system performance by using an efficient and lightweight monitor on all branches, while performing an in-depth analysis only when exceptional program behavior occurs. During the first stage, the hardware monitor stores a trace of all indirect/direct branches inside the branch history buffer (BHB). The second stage then uses the BHB to enforce a number of rules, such as: (1) returns should always point to call-preceding instructions, (2) indirect calls target only function entries, and (3) the number of instructions between indirect branches are usually large. A hardware-based performance evaluation showed a performance overhead of <1.5% for $\gamma$>4 and $\delta$<3. The major hardware costs are: a 32-entry BHB FIFO (which stores the source address, target address, and branch type for each entry), an FSM, and three counters. This design was implemented using the debug interface, and is described in Section 7.1.

## 6.7 Monitoring graph (MG)

Mao et al. [42] proposed to derive an information stream from the executing program to detect attacks. The stream is derived from a combination of any of the following: the address pattern, the opcode pattern, the load/store pattern, the control flow pattern, or a hash of the opcode and instruction address. The expected program behavior, called a monitoring graph, is generated through analysis and simulation, and is then stored in memory. At runtime, the derived stream is compared to the monitoring graph. If there is a mismatch, an attack is assumed and the processor is interrupted. The architecture can detect a stack attack in one to ten cycles, depending on which information is used to derive the stream. The authors reported a monitoring graph size of 100 kB compared to an application binary size of 5 MB. SI can be guaranteed when the derived stream is based on the opcode pattern. The design consists of a watchdog processor connected to the target processor via an unspecified interface.

### 6.8 Branch Regulation (BR)

*Branch Regulation (BR)* [32] protects forward edges by enforcing a simple invariant rule for branch targets, by disallowing arbitrary branches between functions. The enforced invariant rule states that, during normal program execution, a `call` always targets a function entry point, while an indirect `jmp` targets either a function entry or an address within the current function. The enforcement of this rule severely limits JOP attacks because the majority of functions lack a dispatcher gadget (see Section 2), which is critical to launch a JOP attack. The advantages of BR is that it requires no CFG, and it's simple, efficient, lightweight, and severely reduces the set of available gadgets.

BR claims to detect all branches, including unintended branches [32]. However, it seems to provide only limited protection against unintended branches, since indirect branches into the current function are never checked. Kayaalp et al. [32] reported that BR reduced the number of exploitable gadgets to 1% of available gadgets in the original binary. This authors also argue that unless the attacker can find gadgets that execute system calls, the damage from any attack is limited to the compromised process. However, the C standard library contains many system calls, and it seems like nothing prevents the attacker from invoking a function which uses a syscall.

A challenge of using BR is to communicate the function boundaries to the hardware monitor. Kayaalp et al. [32] proposed to communicate each function's bounds with an annotation placed at the first address of each function header. Each `call` instruction can only target annotated addresses, while `jmp` instructions can target either annotated addresses or addresses inside the current function. After branching to the annotated address, the instruction pipeline processes the annotation and places the new function bounds inside a Function Bounds Stack (FBS). Lee et al. [36, 38] proposed to transform each function to start with annotated code which communicates the function bounds by writing to memory-mapped registers (MMIO) (see Section 7).

Kayaalp et al. [32] integrated BR into the execution pipeline stage of a processor. The major hardware components are an SCS, memory for storing the Function Bounds Stack (FBS), and logic for interpreting the annotations and detecting invalid branches. The performance evaluation was done in simulation using PTLsim/x86, and they found that an FBS of 8 entries (total size 96 bytes) leads to a performance loss of about 1%.

### 6.9 Branch Regulation on Basic Blocks (BR on BB)

BB-CFI [18] limits branch targets to basic block boundaries, instead of function boundaries (as in BR [32]). Specifically, branches are only allowed target the first instruction inside a basic block. In addition, a `call` is only allowed to target the first basic block of a function. The authors argue that restricting `jmp` targets to the current function boundary (as was done in BR) is too restrictive, as it cannot support `longjmp()`. A CFG is used to generate metadata containing the start address of each basic block as well as the start of each function. BB-CFI protects against unintended branches, since control can only flow to the first instruction in each basic block. Unlike BR, BB-CFI does not assume DEP. However, the metadata needs to be stored in protected memory. For the evaluation, the authors stored the metadata in hardware on content addressable memory, but it can also be protected by storing it in kernel space, which would incur additional overhead. The authors found that, on average, >99% of gadgets were eliminated from various different benchmarks. The RIPE benchmark, which performs exploits based on code injection, return-into-libc and ROP (a large number of different CRA attacks), was used to evaluate BB-CFI. The authors found that all attacks were blocked (both when $W \oplus X$ is enabled and disabled). It remains to be seen whether a specifically crafted attack can circumvent this security policy.

BB-CFI is integrated into the processor's commit pipeline stage. The major hardware components of the design are an SCS module, the Basic Block Table (BBT) (which contains the profiling metadata), a control unit (CU), and a buffer containing control flow instructions that still need to be validated (Target Address Buffer (TAB)). The BBT is implemented as CAM, which can search for a target address in the entire BBT in a single cycle, and has a maximum memory requirement of 209 kB for one of the measured SPEC benchmarks. A TAB of 1 kB has an estimated a performance overhead of <1%.

## 6.10    Branch Limitation (BL)

Recent works [28, 30] proposed to limit branch targets to basic block entries. This restriction ensures that control can only flow from the exit (last instruction) of one basic block to the first instruction of a basic block. When a branch targets the middle of a basic block it violates the basic block definition, implying that the system is under attack. The BL architectures discussed here only checks forward indirect branches, but can be combined with an SCS to check backward edges. The advantage of this approach is that it requires no CFG.

A major challenge is to communicate the first address of each basic block with the hardware monitor. BBB-CFI [28] argues that confirming a basic block's exit is equivalent to confirming an entry point of the following basic block. Since a branch can only be placed on the last instruction of a basic block, the following instruction is usually the entry point of another basic block. As such, they proposed to restrict indirect branch targets to addresses which are preceded by any branch instruction, including direct/indirect calls, jumps and returns. This ensures that indirect branches can only target the first instruction of a basic block. To support basic blocks which are not preceded by branches, such as fall-through edges of switch-like statements, a second rule is introduced. Here, the the fall-through addresses are stored in metadata (inside main memory). At runtime, if the first rule fails, then the second rule is evaluated. The enforcement of both rules requires performing additional reads: rule one requires fetching the instruction preceding the branch target, while rule two requires reading metadata from memory. The authors reported a 90% reduction in JOP gadgets, and a total gadget reduction of 98.68% when also enforcing an SCS with extended rules. An RIPE evaluation showed that the only successful attacks were return-into-libc attacks which hijack an indirect call to invoke an exported library function. These attacks avoid detection because the call targets a function entry, which seems like a normal control flow to the CFI architecture. The authors reported an average performance overhead of 0.1%.

BBB-CFI [28] is integrated into the processor's pipeline stages. The major hardware components are a 32 element SCS, a buffer containing the most recently verified basic block boundaries, and control logic to compare and fetch metadata from main memory. The fall-through metadata are stored in main memory, with a storage overhead of around 13% when compared to the original program size.

Intel Control-flow Enforcement Technology (CET) [30] solves the challenge of communicating the basic block entry points with the hardware monitor with *indirect branch tracking (IBT)*. Here a new instruction, called ENDBRANCH, is placed at the entry of each basic block that can be invoked through an indirect forward branch. When an indirect forward branch occurs, the following instruction is expected to be an ENDBRANCH, otherwise an attack is assumed. Since this approach is so similar to BBB-CFI, which also enforces the semantics of basic blocks, it is expected to lead to the same gadget reduction count as BBB-CFI. Intel CET is integrated into the pipeline stages of the processor. The major hardware components are an SCS module and a small IBT FSM.

### 6.11 Instruction Set Randomization (ISR)

SOFIA [21] performs Instruction Set Randomization (ISR), where each instruction's bytes are encrypted using control flow information from a CFG, to enforce a fine-grained CFI policy. At runtime, each instruction's bytes are decrypted using a combination of the current and previous program counters. Therefore, any invalid control flow between two instructions will lead to a decryption error. To detect decryption errors and also provide SI, a Message Authentication Code (MAC) is calculated over the instruction bytes in each basic block. A unique key is used for each device, which can only be accessed by the cryptographic hardware, and is only known by the software provider. It is therefore impossible for an attacker to inject code or tamper with the software since he doesn't know the key. SOFIA checks all branches, including unintended branches and fault attacks on the control flow. SOFIA enforces a static CFI policy (no SCS). However, it would be difficult to craft a ROP attack, since all software is stored encrypted, making it near impossible to identify gadgets. SOFIA has an average performance overhead of 106%.

An interesting feature of SOFIA is that it does not only check the validity of control flow between basic blocks, but also verifies the control flow between every two instructions. This is important in scenarios where you need protection against fault attacks on the control flow, or where you cannot rely on the OS to protect the integrity of instructions.

SOFIA is designed for security-critical scenarios, since it prevents the execution of all tampered instructions and all instructions resulting from tampered control flow. This means that it not a single bad instruction will ever execute. This is a useful property in a safety-critical application, where the execution of a tampered instruction can cause a catastrophic failure, e.g., in an insulin pump one bad store instruction can cause the emptying of a patient's insulin tank.

SOFIA [21] is integrated into a processor's pipeline stages. The major hardware components are a MAC component (which consists of a two cycle block cipher), as well as control logic. The MAC component has a critical path which is longer than that of the processor, leading to a clock speed reduction of 23.3%.

### 6.12 Signature Modeling (SM)

*Signature modeling* is a technique used in fault-tolerant computing to detect control flow violations. A checksum is periodically calculated on the executed instructions for comparison with stored reference values. In its simplest form, the comparison is done at the end of each basic block. The reference values are computed offline and stored inside the binary.

In Continuous Signature Monitoring (CSM) [54], a signature of each executing instruction is verified at runtime. The signature is updated (e.g., with an XOR) and verified before the execution of each instruction, which facilitates the enforcement of SI. The signature depends on both the current instruction, as well as previously executed instructions, and it also captures errors in control flow, since a tampered control flow change will eventually lead to different instructions being executed. To ensure that the signature size does not become too large, only a small amount of signature bits are stored and checked for each instruction. Since the error propagates within the signature register, almost all control flow errors can be detected within 3 checks for a signature check size of only 4 bits. The advantage of this approach is that it can detect control flow changes at a fine granularity, such as when a fault attack is used to skip a single instruction inside a basic block. The disadvantage is that a signature value for every instruction is necessary.

*Generalized Path Signature Analysis (GPSA)* [41] relies on signature updates in the code to ensure that a signature value at a given location is always the same, no matter which (valid) path was taken through the CFG. First, the CFG is divided into a number of path sets, with each path set starting and ending at the same node. The goal is then to have a single signature value for each path set no matter which control flow path was taken. To achieve this, justifying signatures are inserted on some of the paths of each path set. When control flow arrives at the end of each path set, the signature is compared to the reference value. This lowers the storage requirement for reference values, as integrity checks occur less frequently. However, this method increases the latency between fault and detection.

Werner et al. [53] used GPSA together with CSM to protect software running on an embedded processor against fault-based control flow attacks. For CSM, the processor's fetch unit was modified to also fetch the reference signature. Only when the reference signature matches the current signature will the instruction be forwarded to the decode stage. A post processing tool was used to obtain a CFG, and calculate the derived signatures. In addition, binary instrumentation was used to send commands to the hardware to update/check the signature (by writing to a memory mapped register). All branches (direct and indirect) are checked, since any control flow change causes an update to the runtime signature. They reported a performance overhead of 9%.

Werner et al. [53] integrated their hardware monitor into a commercial ARM Cortex-M3 processor. The signature monitor was implemented as a memory mapped device. The major hardware components are a CRC-based signature monitor and logic to fetch reference values from main memory and compare them against the derived signatures.

### 6.13   Code Pointer Integrity (CPI)

Code Pointer Integrity (CPI) policies aim to prevent control flow hijacking by protecting the integrity of code pointers at runtime.

A recent whitepaper [2] discusses hardware-based CPI support on the new ARMv8.3-a ISA. Here, short MAC tags, called Pointer Authentication Codes (PACs), are used to verify the integrity of pointers at runtime. The MAC is calculated over the pointer target together with a context, which is usually the pointer address, e.g., PAC = $MAC_k$(target,context). Whenever a code pointer is used, it's integrity is first verified, and whenever a pointer target is updated, a new PAC is calculated and stored. This prevents pointer tampering, since an attacker will also has to forge the PAC. Relocating a PAC and pointer pair to a new address is infeasible, since the MAC computation includes the pointer address. Key management is done by privileged software (EL1, EL2, EL3), and it is expected that the higher privilege levels control the keys for the lower privilege levels, which includes assigning a unique key per each process or per each boot.

PAC exploits the fact that the available virtual address space in 64-bit architectures is less than 64-bits, e.g., ARM64 Linux uses a 40-bit address space by default, leaving 26 bits to be used for storing PAC values. The PAC size can vary between 3 to 31 bits, depending on the system's virtual address space configuration, and the PAC is placed in the unused upper bits of the pointer before being written to memory.

The core functionality of PAC is provided by two instructions types: PAC* computes and stores a PAC, while AUT* verifies a PAC and restores the pointer value. An integrity failure during AUT* leads to a pointer update to an illegal address, which causes an exception when the invalid pointer is dereferenced. The idea is that the compiler inserts these instructions inside the binary in order to update and check the integrity of the critical pointers at runtime. PAC is integrated into the pipeline stages of the processor, since it uses dedicated instructions. The major hardware components are a low latency MAC implementation which uses the QARMA block cipher, together with control logic to calculate and verify PACs.

It is important to note that PAC is susceptible to pointer substitution attacks, where one authenticated pointer is replaced with another. Since the MAC computation includes the pointer address, this will only work if the pointer is replaced with an authenticated pointer that was previously located at the same address.

Software-based CPI [35, 43, 52] enforces pointer integrity by storing sensitive pointers in an isolated memory region, and further uses runtime checks to verify the correctness of each code pointer on each control transfer. Software-based CPI which rely on information hiding have been demonstrated to be broken [23]. However, other instruction-level isolation mechanisms, such as software fault isolation [35], provide much stronger protection for storing pointers, and have not been demonstrated to be broken. PAC [2] is unlikely to be susceptible to these attacks, since each stored pointer is protected by a cryptographic MAC, which makes tampering difficult.

## 7  CFI ENFORCEMENT VIA THE DEBUG INTERFACE

A common approach to enforce CFI through the debug port is to configure the debugger to provide a trace for each control flow transfer. However, the traces often do not provide all the required branching information to enforce a CFI policy. A simple solution is to modify the debugger to include the missing information inside the debug trace ([27, 28, 36]). However, this is not ideal, since the main reason for using the debug interface is to avoid having to modify existing IP (see Section 3.3.2). Another approach is to provide supplementary metadata inside main memory [37], or to communicate the missing information with the hardware by writing to memory-mapped addresses [38, 39].

### 7.1  Implementations

Lee et al. [37] observed that the debug traces generated by an ARM CoreSight debugger contains only the branch target address for indirect branches, while for direct branches only the direction (taken/not taken) is available. In order to enforce an SCS, the branch type and source address are also necessary. To this end, offline analysis on the binary is used to generate supplementary metadata. The metadata contains the branch type, source address, and target address for each branch instruction (at the end of each basic block). By observing the debug traces, the hardware monitor can determine which basic block is currently executing, thereby allowing accurate enforcement of the SCS policy. The metadata increased the binary size by 145.5%, while incurring a performance penalty of 2.4%, which is mostly due to bus contention since the main memory is shared between the monitor and the processor. The major hardware components are a 16 entry branch trace FIFO, a debug trace analyzer, logic to read metadata from main memory, and an SCS module. The hardware monitor was implemented in reconfigurable fabric, which runs at a lower clock frequency than the ASIC processor. They reported an operating frequency of 90 MHz for the ROP monitor, and 200 MHz for the host CPU.

Lee et al. [38, 39] proposed to supplement ARM CoreSight's trace information by exploiting the fact that the target address is only available for indirect branches. To do this, a trampoline is inserted for each `call` in the binary, together with replacing each `call` instruction with an indirect jump to the unique trampoline address associated with that call. Since each trampoline has a unique address, the branch type and source address can be derived by checking the target address of the direct jump. This approach has a code size increase of 16.6%. The major hardware components are a 32-entry branch trace FIFO, a debug trace decoder logic, a 32-entry MMIO FIFO, a branch trace analyzer, a FIFO trace combiner, a CRA detector controller, and an SCS module. The branch trace FIFO acts as a buffer for receiving debug traces, while the MMIO FIFO acts as a buffer for receiving the function bounds via MMIO. On the Zynq platform, the CRA monitor runs at 60 MHz, while the ARM processor runs at 150 MHz. They reported a performance overhead of 3%, which is mostly due to memory contention.

Lee et al. [36] enforced BR and an SCS through the Core Debug Interface (CDI). The debug interface was modified to provide more information for the hardware monitor, and Memory-Mapped IO (MMIO) is used to receive function boundary information from the instrumented binary. They reported an average performance overhead of less than 2%.

The major hardware components are a debug trace filter, an indirect branch bounds checker, an internal bus, main control logic, the memory region protector, a secure call stack, and a debug trace FIFO.

CONVERSE [27] enforced a table-based policy through a Commercial off-the-shelf (COTS) watchdog processor that is connected to the target processor's Nexus 5001 debug port. Each debug trace contains the source and target addresses. When the watchdog processor detects of invalid control flow, the debug BREAK feature is used to halt execution of the target processor. The authors reported that their implementation does not have any overhead.

BBB-CFI [28] enforced an SCS and BL by exploiting the Intel's Last Branch Register (LBR) and Processor Trace (PT) features. The LBR buffers the 16 most recent traces, with each trace consisting of the source and target address, while the PT is used to write out a trace to physical memory or to a dedicated port on a SoC. BBB-CFI modified the LBR to include a 2-bit branch type field.

### 7.2 Limitations

A common problem with enforcing a CFI policy via the hardware debugger is that the hardware monitor could drop traces given a sufficiently high branch rate [27, 39]. This happens when the hardware monitor requires more time to process a trace than the rate at which branches occur on the target processor. Whenever a trace is dropped without being analyzed, it introduces a security weakness, since an attacker could exploit this by launching an attack during a time period when branches are frequently occurring. This problem can be mitigated by using a FIFO to store all incoming debug traces before they are processed. However, given a sufficient number of branches over a short time period, the hardware monitor will eventually fall behind, leading to dropped traces. This is especially problematic if the hardware monitor operates at a much lower clock rate than the processor.

Another problem is that the hardware debugger can be used by an attacker to circumvent the security of the system. If the attacker can access the debug interface, he could use it to tamper with code and data memory, or even disable the hardware monitor by tampering with the tracing mechanism. Therefore, care needs to be taken to ensure that an attacker cannot obtain access to the debug interface. This could be done by ensuring that no external debug interface is present, and prohibiting software running on the processor from accessing the debugger.

## 8 COMPARISON OF ARCHITECTURES

Table 1 shows a comparison of all hardware-based CFI architectures analyzed in this paper. In general, most architectures enforce dynamic CFI which relies on an SCS for stateful backward edge protection, while using a range of different static forward-edge policies.

### 8.1 Protection provided

Most works assume attackers in control of data memory, while a smaller number of architectures assume an attacker that can control code memory or perform fault attacks on control flow. Whenever an attacker model wasn't specified, we made an estimate of the assumed attacker, which is indicated by "*". The following text discusses the protection provided in terms of the assumed attacker model.

A policy which prevents arbitrary branches imposes limits on the allowed branch targets of forward branch instructions. We use the "indirect branch protection" and "direct branch protection" columns to indicate whether an architecture can prevent arbitrary branches to respectively for indirect or direct instructions. Many architectures can prevent arbitrary indirect branches, while only some can prevent arbitrary direct branches. Indirect branch protection

Table 1. Overview of hardware-based CFI architectures.

| Architecture | Policies | | Protection Provided | | | | | | | Requirements | | | | | |
| --- | --- | --- | --- | --- | --- | --- | --- | --- | --- | --- | --- | --- | --- | --- | --- |
| | Stateful policy | Static policy | Attacker Capabilities[1] | Indirect Branch[3] | Direct Branch[3] | Data structures[2] | Exploitable Gadgets[4] (%) | Software Integrity | Fine-grained | CFG | SW in TCB | ISA changes | Debug Port | W⊕X | Academic |
| SmashGuard [45] | SCS | ○ | D | ◐ | ◐ | ● | − | ○ | ○ | ○ | ○ | ● | ○ | ○ | ● |
| IBMAC [24] | SCS | ○ | D | ◐ | ○ | ● | − | ○ | ○ | ○ | ○ | ● | ○ | ○ | ● |
| Lee et al. [37] | SCS | ○ | D | ◐ | ○ | ◐ | − | ○ | ○ | ○ | ○ | ● | ● | ○ | ● |
| HAFIX [19] | ● | ○ | D | ◐ | ○ | ↯ | − | ○ | ○ | ● | ● | ● | ○ | ○ | ● |
| HCFI [15] | SCS | Label | D | ● | ● | ● | 0* | ○ | ● | ● | ● | ● | ○ | ○ | ● |
| Sullivan et al. [51] | SCS | Label | D | ● | ● | ● | 0* | ○ | ● | ● | ● | ● | ● | ○ | ● |
| Arora et al. [6] | SCS | Table + FSM | CDP* | ● | ● | ◐ | 0* | ● | ● | ● | ○ | ● | ○ | ○ | ● |
| CONVERSE [27] | ○ | Table | − | ● | ● | ● | 0* | ● | ● | ● | ● | ● | ○ | ○ | ● |
| Rahmatian et al. [46] | ○ | FSM | CD | ◐ | ○ | ○ | 100* | ○ | ○ | ● | ● | ● | ○ | ○ | ● |
| SCRAP [33] | SCS | Heuristic | C* | ● | ○ | ○ | − | ○ | ○ | ○ | ○ | ● | ○ | ○ | ● |
| Lee et al. [39] | ○ | Heuristic | D | ◐ | ◐ | ● | − | ○ | ○ | ● | ● | ● | ● | ○ | ● |
| Mao et al. [42] | SCS | MG | CDP* | ● | ● | ○ | 0* | ● | ● | ● | ○ | ● | ○ | ○ | ● |
| Werner et al. [53] | SM | SM | CDP* | ● | ● | ○ | 0* | ● | ● | ● | ● | ● | ○ | ○ | ● |
| BR [31, 32] | SCS | BR | D | ◐ | ● | ● | 1 | ○ | ○ | ● | ○ | ● | ○ | ○ | ● |
| Lee et al. [38, 39] | SCS | BR | D | ◐ | ● | ● | 1* | ○ | ○ | ● | ○ | ● | ● | ○ | ● |
| Lee et al. [36] | SCS | BR | D* | ◐ | ● | ● | 1* | ○ | ○ | ● | ● | ● | ● | ○ | ● |
| BB-CFI [18] | SCS | BR on BB | CD | ● | ● | ◐ | 1 | ○ | ○ | ● | ○ | ● | ○ | ● | ● |
| BBB-CFI [28] | SCS | BL | D | ○ | ○ | ● | 1.3 | ○ | ○ | ○ | ○ | − | ● | ● | ● |
| Intel CET [30] | SCS | BL | D* | ○ | ○ | ● | 1.3* | ○ | ○ | ○ | ○ | ● | ○ | ● | ○ |
| SOFIA [21] | ○ | ISR | CDP | ● | ● | ○ | 0* | ● | ● | ○ | ● | ● | ○ | ● | ● |
| ARMv8.3-a [2] | CPI | CPI | D | ● | ● | ○ | 0* | ● | ● | ○ | ● | ● | ● | ○ | ○ |

● = Yes; ◐ = Partial; ○ = No; ↯ = Not Applicable; − =Unspecified; * = Estimated
[1]D = Data modifications; C = Code modifications; P = Physical attacks
[2]Metadata, SCS, or runtime data structures access protection.
[3]Prevent forward branches to arbitrary target addresses; [4]Exploitable Gadgets in the binary

is important to protect against JOP and ROP attacks, since these attacks make use of gadgets consisting of indirect branches. Direct branch protection provides an additional protection layer, and is important in the following two scenarios. First, when the attacker controls code memory direct branch targets could be modified, since code is mutable. Second, when fault attacks on the control flow occur, direct branch targets could be tampered with. BR provides only partial indirect branch protection, since indirect branches to any address within the current function are allowed. This leaves BR somewhat vulnerable to unintended branches, since it allows CRAs which do not cross function boundaries.

Many solutions store sensitive runtime data structures or metadata in data memory. However, less than half the architectures protect both their runtime (SCS) data and metadata which is stored in main memory. Since all attacker can control data memory, it is important to protect sensitive data stored in main memory in order to prevent attackers from circumventing the security policy by tampering with the stored data.

Coarse-grained policies cannot prevent all indirect branches from disobeying the CFG. As such, their program binaries contain a number of gadgets which are usable by an attacker. To quantify the number of usable gadgets, a metric called gadget reduction or Average Indirect target Reduction (AIR) is sometimes reported, which we summarized in the "*Exploitable gadgets*" column. The metric reports the ratio between the number of reachable gadgets when CFI is enforced, compared to the reachable gadgets when CFI is not enforced (i.e all gadgets). Gadgets are identified under the assumption that each gadget ends with an indirect branch and has no side effects, which can be automated with a gadget finder tool, such as ROPGadget [47]. It is important to remember that a successful CRA attack requires only a handful of gadgets. Therefore, even with large gadget reductions, attacks can still succeed, albeit with more effort from the attacker. In some cases the values reported in the "Exploitable gadgets" column are estimated from the reported values of other architectures with the same policies. e.g., the values for [30] were based on [28] which both enforced BL, while [36, 38] are based on [32], since they enforce the same policies. Heuristic-based approaches [33] cannot prevent branches, since they only detect abnormal branch behavior, and therefore almost all gadgets are exploitable. Fine-grained policies prevent arbitrary branches, and therefore the exploitable gadgets are estimated to be close to zero.

Most works assume attackers in control of data memory, while not controlling code memory, since it is protected with page-based protection, such as W ⊕ X . In addition, these works only protect indirect forward branches. However, architectures which assume attackers which also control code memory, typically prevent code tampering/injection through an SI mechanism, which verifies code integrity at runtime by means of a cryptographic hash/MAC.

## 8.2 Requirements

This subsection compares the architectural requirements to the protection provided.

We found that roughly a third of the architectures offer fine-grained protection which relies on a precise CFG, as indicated by the "*CFG required*" column. Fine-grained policies supposedly provide stronger security guarantees, while the security of the policy depends on the accuracy of the CFG. However, their dependence on a precise CFG is an important limitation, since it is difficult to obtain a precise CFG from complex binaries (see Section 5.3.1). We found that most architectures which do not rely on a precise CFG are coarse-grained, and therefore have a non-zero amount of exploitable gadgets. The exception seems to be ARMv8.3-a which is the only architecture which provides fine-grained protection while not relying on a CFG. However, currently we cannot thoroughly evaluate the architecture, since official reference material has not been released yet.

More than half of the architectures rely on software placed inside the binary to implement its security policy, as indicated by the "*SW in TCB*" column. This is a reasonable requirement, provided that code integrity is guaranteed, e.g., through W ⊕ X or by enforcing SI. All architectures respected this requirement.

We found that most architectures are integrated into the instruction pipeline stages of the processor, as indicated by the "*ISA changes*" column. In contrast, some architectures exploit the debug interface to allow enforcing their policies, as indicated by the "*Debug Port*" column. The main advantage of using the debug port is that it allows for enforcing a policy without making changes to the processor or other existing IP. However, we found that some architectures enforced via the debug port also modified the ISA to enforce their policies.

We found only two non-academic architectures, namely Intel CET [30] and ARMv8.3-a [2], as indicated by the "*Academic*" column. At the time of writing, Qualcomm has only released a whitepaper describing the PAC architecture, while Intel has released a detailed "Technology Preview" document. Intel's solution, which performs a type of Branch Limitation (BL), is coarse-grained, but practical for widespread deployment, since it only requires the insertion of special landing pad instructions. ARM's solution, which relies on CPI, promises to offer fine-grained protection, and seems to be practical since it does not require complex binary transformations. Both solutions do not rely CFGs, which make these architectures practical for widespread deployment.

### 8.3 Overhead

Table 2 compares the performance and area overhead of the CFI architectures which were implemented on a Field Programmable Gate Array (FPGA) or Application-Specific Integrated Circuit (ASIC). The architectures which have no published performance or hardware overheads at the time of writing were not listed in the table, namely Mao et al. [42], Intel CET [30], and ARMv8.3-a [2].

Szekeres et al. [52] found that security mechanisms with a performance overhead of more than 10% do not gain widespread adoption in production environments. In addition, many people believe that the average performance overhead should be less than 5% in order to obtain adoption from industry. The *"Performance Overhead"* column shows that most architectures fall below the 5% barrier.

The majority of the architectures were evaluated for both performance and area in hardware. However, some architectures [31, 32, 45] were only evaluated in simulation, while some others ([6, 18, 28, 33]) performed a simulation-based performance evaluation, and made a separate a hardware implementation to evaluate the hardware area overhead.

Many architectures performed an evaluation on FPGA, but failed to specify the target FPGA technology, which makes it impossible to compare their hardware overhead with other designs. Some designs report their area (LUTs, registers, BRAM, gates) as an increase in area compared to the original unmodified processor. This also makes it difficult to compare the overhead with other designs, since the area used by different processors differ greatly. In addition, large variations in area can be observed for the same processor under different configurations. Many architectures are integrated into the pipeline stages of the processor, and it is important to know if the design has an impact on the critical path of the processor. Therefore, it is best to report the maximum attainable clock speed of the modified processor compared to the stock processor.

## 9 CONCLUSION

In this paper, we presented an analysis of the security policies used by 21 hardware-based CFI architectures. This included a detailed comparison of the used policies with respect to their security, limitations, hardware cost, performance, and practicality. We found that most architectures protect backward edges with a Shadow Call Stack (SCS), and a large body of work discusses the intricacies of enforcing an SCS. However, for forward edge protection we found that a range of different policies are used, with differences in their security and practical limitations.

In general we found that a lot of progress has been made in the area of CFI. In particular, SCS provides excellent protection against Return-Oriented Programming (ROP), and does not suffer from any practical limitations or security problems which prevents widespread adoption. However, static forward-edge CFI policies face practical limitations, such as:

Table 2. Performance and hardware overheads of the CFI architectures. All reported percentages are relative to the baseline performance of the target processor, while the non-percentages are absolute values.

| Architecture | Static policy | Target ISA | Perf. Overhead (%) | Simulation† | Technology | LUTs | Flip Flops | 36K BRAM | Area (%) | Clock (MHz) | Gates (kGE) | Technology (nm) | Area (%) | Clock (GHz) |
|---|---|---|---|---|---|---|---|---|---|---|---|---|---|---|
| | | | **Perf. Evaluation** | | **FPGA** | | | | | | **ASIC** | | | |
| SmashGuard [45] | ○ | Alpha | 2.8% | ● | – | – | – | – | – | – | – | – | – | – |
| IBMAC [24] | ○ | ATMega103 | – | ○ | Cyclone-2 | 215 | 0 | 0 | 9% | – | – | – | – | – |
| Lee et al. [37] | ○ | Cortex-A9 | 2.4% | ○ | Zynq | 7362 | – | 5 | – | 90 | 86.7 | 45 | – | – |
| HAFIX [19] | ○ | LEON3 | 2% | ○ | – | 0.3% | 3% | 8% * | – | 0% | – | – | – | – |
| HAFIX [19] | ○ | Siskiyou Peak | 2% | ○ | – | <1% | 2.5% | 2 | – | – | – | – | – | – |
| HCFI [15] | Label | LEON3 | 1% | ○ | – | 2.6% | 2.5% | – | – | 0% | – | – | – | 0% |
| Sullivan et al. [51] | Label | LEON3 | 1.8% | ○ | – | – | – | – | – | – | 32/28 | 1.78% | – | 3 |
| Arora et al. [6] | Table + FSM | ARM920T | – | ● | Virtex-2P | 1839 | – | – | – | 57 | – | 130 | 9.07% | – |
| CONVERSE [27] | Table | – | 0% | ○ | – | – | – | – | – | – | – | – | – | – |
| Rahmatian et al. [46] | FSM | LEON3 | 0% | ○ | Virtex-6 | 340 | – | 2 | 1.9% | 0% | – | – | – | – |
| SCRAP [33] | Heuristic | x86 | 2% | ● | Spartan-3 | – | – | – | – | 284 | – | – | – | – |
| Lee et al. [39] | Heuristic | Cortex-A9 | 1.5% | ○ | Zynq | 1795 | – | 2 | – | 60 | – | – | – | – |
| Werner et al. [53] | SM | Cortex-M3 | 9% | ○ | – | – | – | – | – | – | 130 | 6.4% | – | – |
| BR [31, 32] | BR | x86 | 2% | ● | – | – | – | – | – | – | – | – | – | – |
| Lee et al. [38, 39] | BR | Cortex-A9 | 4.5% | ○ | Zynq | 3172 | – | 3 | – | 60 | – | – | – | – |
| Lee et al. [36] | BR | LEON3 | 2% | ○ | Virtex-5 | 22.7% | – | 15% | – | – | 244.2 | 45 | 13.79% | 1 |
| BB-CFI [18] | BR on BB | x86 | <1% | ○ | Virtex-5 | 86 | – | 4 | – | 313 | – | 65 | 0.02% | – |
| BBB-CFI [28] | BL | x86 | 0.1% | ○ | Virtex-6 | 3720 | 3404 | – | – | 231 | – | – | – | – |
| Intel CET [30] | BL | x86 | – | ○ | – | – | – | – | – | – | – | – | – | – |
| SOFIA [21] | ISR | LEON3 | 106% | ○ | Virtex-6 | 958 | 467 | 0 | – | 23.2% | – | – | – | – |

† Performance evaluation done in simulation, and not on the hardware of the target ISA
– =Unspecified; * = distributed RAM

- The security of fine-grained CFI relies on the precision of a Control Flow Graph (CFG), which cannot be accurately generated for binaries containing indirect forward branches. ARMv8.3-A seems to provide fine-grained protection without the need for a CFG. However, it has not been publicly released yet, which means that the security community still has to evaluate it's limitations.

- Coarse-grained CFI relies on a relaxation in the strictness of the policy by enforcing simple rules that do not require a CFG. However, this comes at a reduction in the security provided, since it cannot detect all illegal branches.

- Heuristics provide coarse-grained and practical protection. However, it can be circumvented by an attack which is crafted to thwart the heuristic. It further suffers from false positives, and proper heuristic parameter selection remains unsolved.

These limitations could explain why industry has not yet publicly released a CFI-capable processor, and we believe that the problem of enforcing practical and fine-grained CFI still remains unsolved.